\documentclass[12pt,a4paper]{article}
\usepackage{graphicx}
\usepackage{amsmath}
\usepackage{amssymb}
\usepackage{bbold}

\begin{document}

\begin{center}
{\LARGE The Planar Thirring
Model\\ with K\"ahler-Dirac Fermions}

\vskip 1cm
{\bf Simon Hands}

\vskip 0.5cm
{\em Department of Physics, \\Faculty of Science \& Engineering, Swansea University,\\
Singleton Park, Swansea SA2 8PP, United Kingdom. }

\end{center}

\begin{abstract}
K\"ahler's geometric approach in which relativistic fermion fields are treated as
differential forms is applied in three spacetime dimensions.  It is shown that
the resulting continuum theory is invariant under global U($N)\otimes$U($N)$ field
transformations, and has a parity-invariant mass term, both symmetries shared
in common with staggered lattice fermions. The formalism is used to construct a
version of the Thirring model with contact interactions  between conserved
Noether currents. Under reasonable assumptions about field rescaling after
quantum corrections, a more general interaction term is derived, sharing the
same symmetries but now including terms which entangle spin and taste degrees of
freedom, which exactly  coincides with the leading terms in the staggered lattice
Thirring model in the long-wavelength limit. Finally truncated versions of the
theory are explored; it is found that excluding scalar and
pseudoscalar components leads to a theory of six-component fermion
fields describing particles with spin 1, with fermion and antifermion
corresponding to
states with definite circular polarisation. In the UV limit only transverse
states 
with just four non-vanishing components propagate. Implications
for the description of dynamics at a strongly interacting renormalisation-group
fixed point are discussed.
\end{abstract}

\section{Introduction}
This paper concerns relativistic fermions interacting strongly in three
spacetime dimensions, in the context of a field theory known as the Thirring
model with Lagrangian density
\begin{equation}
{\cal L}
=\bar\psi_i(\partial\!\!\!/\,+m)\psi_i+{g^2\over2N}(\bar\psi_i\gamma_\mu\psi_i)^2.
\label{eq:contThir}
\end{equation}
Here the fields $\psi_i,\bar\psi_i$ are reducible spinors, so the Dirac
matrices $\gamma_\mu$ are $4\times4$. In Euclidean metric they obey
$\gamma_\mu=\gamma_\mu^\dagger$ and 
$\{\gamma_\mu,\gamma_\nu\}=2\delta_{\mu\nu}$. The index $i=1,\dots,N$ runs over
$N$ distinct fermion species. The contact interaction between conserved
fermion currents $\bar\psi\gamma_\mu\psi$ results in a repulsive force between
fermions, but attraction between fermion and antifermion. A question of interest is then
whether in the massless limit $m\to0$ a bilinear condensate
$\langle\bar\psi\psi\rangle\not=0$ forms as a result of strong interactions,
leading to the dynamical generation of fermion mass.

It is natural to analyse bilinear condensation in terms of symmetry breaking.
In three dimensions there are two elements of the reducible Dirac algebra $\gamma_4$
and $\gamma_5$ which anticommute with the kinetic term of (\ref{eq:contThir}).
Accordingly for $m\to0$ (\ref{eq:contThir}) is invariant under the following
field rotations:
\begin{eqnarray}
\psi\mapsto e^{i\alpha_1}\psi;\;\bar\psi\mapsto\bar\psi
e^{-i\alpha_1}\;\;&:&
\psi\mapsto e^{\alpha_{45}\gamma_4\gamma_5}\psi;\;\bar\psi\mapsto\bar\psi
e^{-\alpha_{45}\gamma_4\gamma_5};\\
\psi\mapsto e^{i\alpha_4\gamma_4}\psi;\;\bar\psi\mapsto\bar\psi
e^{i\alpha_4\gamma_4}\;\;&:&
\psi\mapsto e^{i\alpha_{5}\gamma_5}\psi;\;\bar\psi\mapsto\bar\psi
e^{i\alpha_{5}\gamma_5}.\label{eq:45}
\end{eqnarray}
Together these rotations generate a U(2$N$) global symmetry which can be broken
either explicitly by $m\not=0$ or spontaneously by $\langle\bar\psi\psi\rangle\not=0$
to U($N)\otimes$U($N$), when the rotations (\ref{eq:45}) no longer leave the
ground state invariant. Goldstone's theorem implies spontaneous symmetry
breaking results in $2N^2$ massless bosons in the theory's spectrum. 

It is suspected that symmetry breaking occurs for sufficiently large
interaction strength $g^2$ and sufficiently small $N$; it is even possible
that the resulting quantum critical point observed at $g_c^2(N)$ might be a
UV-stable fixed point of the renormalisation group, implying that a continuum 
limit at this point is possible.  The fixed-point theory is expected to display universal
features of the strongly-interacting dynamics characterised by the
pattern of symmetry breaking. However, there are no
small parameters to enable a systematic investigation of this phenomenon by
analytic means. Determination of the critical exponents, and even the critical
flavor number $N_c$ below which symmetry breaking can occur, are essentially
non-perturbative problems. 

A natural approach employs numerical simulations of
lattice field theory (a recent review can be found in \cite{suranyi}). Most
recent work uses a lattice fermion formulation which seeks to respect the U($2N$)
symmetry such as the SLAC derivative~\cite{Wellegehausen:2017goy,Lenz:2019qwu}, 
or domain wall fermions~\cite{Hands:2018vrd,Hands:2020itv}. However, there is
also a substantial body of earlier simulations~\cite{DelDebbio:1997dv} employing the more primitive
staggered formulation, in which fermion fields are represented by
single-component Grassmann objects $\chi,\bar\chi$ located on the sites of a
cubic lattice. As well as U($N$) flavor rotations, staggered fermions also enjoy
a second U($N$) global symmetry protecting them
from acquiring mass, of the form
\begin{equation}
\chi\mapsto e^{i\beta\varepsilon_x}\chi;\;\bar\chi\mapsto
e^{i\beta\varepsilon_x}\bar\chi,
\label{eq:u1epsilon}
\end{equation}
where $\varepsilon_x=(-1)^{x_1+x_2+x_3}$ is an alternating sign in effect partitioning the
sites $x$  into distinct odd and even sublattices.  
This time, therefore, bilinear condensation
drives a symmetry breaking U($N)\otimes$U($N)\to$U($N$), resulting in just $N^2$
Goldstones. For a strongly-interacting system, therefore,  we expect
distinct fixed-point behaviour, and indeed simulations of the staggered
model~\cite{Christofi:2007ye} support a
critical $N_c\approx3.3$ significantly larger than that found for the U($2N$)-symmetric
variants~\cite{Lenz:2019qwu,Hands:2018vrd,Hands:2020itv}. Moreover,
simulation studies of the minimal staggered model with
$N=1$\footnote{It is very common in the literature to designate $N$
staggered flavors in 3$d$ as describing $N_f=2N$ ``continuum
flavors''~\cite{Burden:1986by}.}
~\cite{Chandrasekharan:2011mn} find critical indices
indistinguishable from those of the Gross-Neveu model having the same global
symmetries~\cite{Chandrasekharan:2013aya}, even though in the latter case symmetry breaking can be
described analytically using a $1/N$ expansion. 

Despite these apparent shortcomings the staggered Thirring model does exhibit
interesting behaviour; in particular, the critical exponents characterising the
fixed-point are particularly sensitive to the value of $N<N_c$. Could there
exist a
continuum-based description of the corresponding fixed-point theories?  One
question which needs addressing is the significance of $N$ -- in a
weak-coupling long-wavelength limit it is natural to interpret staggered femions
in terms of $N_f=2N$ autonomous flavors~\cite{Burden:1986by}, or in modern
parlance, 
each staggered flavor describes two continuum ``tastes''. However, even in
early staggered Thirring studies~\cite{DelDebbio:1997dv} the factorisation of interaction
currents into mutually-distinct taste sectors was not manifest, and there is no reason
{\em a priori\/} to require this in a strongly-coupled setting. In what follows
we will refer to the difficulty in separating taste and spin components as
``spin/taste entanglement''. A related question is how to engineer the
U($N)\otimes$U($N$) symmetry in the continuum where we have no lattice partition
to help recover (\ref{eq:u1epsilon}).

This paper will answer such questions using a framework introduced into lattice
field theory by Becher and Joos in 1982~\cite{Becher:1982ud}, who found that a
version of the Dirac equation rooted in concepts of differential geometry
originally noted by K\"ahler~\cite{Kaehler}  in 1962 is in fact the formal continuum limit of
staggered lattice fermions. As set out in the next few sections, in the K\"ahler-Dirac
approach fermions are not spinor fields but rather are complexes of $p$-forms,
where $p=0,1,\ldots,d$, with $d$ the dimension of spacetime. This is a natural 
way to prepare for transcribing continuum fields to a
lattice~\cite{Rabin:1981qj}.
Each $p$-form has
$_dC_p$ components. In the four
dimensional case analysed in \cite{Becher:1982ud} a fermion field has thus
$1+4+6+4+1=16$ components, which are recast as 4 independent tastes of
4-component spinor fields. For the case $d=3$ to be developed in what follows,
the corresponding field has 8 components recast as 2 tastes of reducible
4-component spinor. The algebraic details very closely mirror the assignment of spin/taste
degrees of freedom to staggered lattice fermions in three spacetime dimensions
originally set out by Burden and
Burkitt~\cite{Burden:1986by}. 

The remainder is organised as follows. Sec.~\ref{sec:prelim} is a brief but
hopefully self-contained introduction to the differential geometry
machinery required. Readers who are already expert will find our notations
and conventions set out; those less familiar might also benefit from the 
helpful Appendix of \cite{Becher:1982ud}, or a textbook such as \cite{GS}.
Sec.~\ref{sec:KDE} derives the equivalence between the free K\"ahler-Dirac equation,
which with suitable notation assumes the same form in any dimension, and a continuum Dirac equation in
three Euclidean dimensions describing two tastes of reducible spinor. The same
framework is used in Sec.~\ref{sec:current}, following the introduction of a
generalised scalar product between $p$-forms,  to identify the fermion current that
will be used in building the Thirring interaction term. Sec.~\ref{sec:action}
at last introduces the Thirring model action in the K\"ahler-Dirac language, and
identifies both the U($N)\otimes$U($N$) global symmetry and also an important  parity
symmetry shared in common with staggered lattice fermions. The Euclidean path
integral is introduced permitting an explicit derivation of the 
Noether current associated with the symmetry corresponding to
(\ref{eq:u1epsilon}). 

In Sec.~\ref{sec:qc} we begin to take the geometrical form
of the theory more seriously by exploring the idea that in a suitably-regularised
interacting theory the renormalisation of the field components should depend on
$p$: the Thirring interaction term is modified in order to accommodate this
possibility, and it is shown that the resulting terms when
recast in a spinor basis exhibit
spin/taste entanglement, and are in exact correspondence with the interaction
derived from the staggered Thirring model~\cite{DelDebbio:1997dv}  using the formalism of
\cite{Burden:1986by}. This demonstrates that the proposed $p$-dependent field
rescaling is perfectly consistent with a properly-regularised lattice model, and
also that spin/taste entanglement is not a lattice artifact but rather in fact a
feature of an interacting continuum field theory.  Finally in Sec.~\ref{sec:trunc} the idea is
taken a step further with the exploration of truncated actions resulting from retaining
just field components with two consecutive values of $p$. The most interesting
case corresponds to keeping just $p=1,2$, resulting in a theory of
six-component spin-1 fermions, whose physical states are transverse, and for
which fermion and antifermion are states of opposite polarisation.
Sec.~\ref{sec:disc} summarises the paper's findings and speculates on the
applicability of the exotic scenario of Sec.~\ref{sec:trunc} to the physics of a
putative renormalisation-group fixed point at strong coupling. 

\section{Mathematical Preliminaries}
\label{sec:prelim}

The theory to be developed uses the language of differential forms in
three-dimensional Euclidean spacetime.  We will follow the presentation and
notation of \cite{Becher:1982ud} closely. In this approach all physical quantities
are viewed as $p$-forms defined in some vector space $^p\Lambda$, with
$p=0,1,\ldots,3$. A suitable basis for $^p\Lambda$ is given by
$dx^H=dx^{\mu_1}\wedge\ldots\wedge dx^{\mu_p}$ with the exterior
product satisfying
\begin{equation}
\wedge:\;^p\Lambda\times^q\Lambda\mapsto\,^{p+q}\Lambda,\;\;\;dx^H\wedge
dx^K=\begin{cases}
\rho_{H,K}dx^{H\cup K} & \mbox{if}\; H\cap K=\emptyset;\\
0 & \mbox{otherwise,}
\end{cases}\label{eq:exterior}
\end{equation}
where the sign factor $\rho_{H,K}=(-1)^s$, with $s$ the number of pairs
$\mu,\nu\in H\times K$ with $\mu>\nu$, and $\rho_{\emptyset,H}=\rho_{H,\emptyset}=1$.
With this in place any function $\Phi$ can be expanded as follows:
\begin{eqnarray}
\Phi(x)&=&\varphi_\emptyset(x)+\varphi_\mu(x)dx^\mu+{1\over{2!}}\varphi_{\mu\nu}(x)dx^\mu\wedge
dx^\nu+\varphi_{123}(x)dx^1\wedge dx^2\wedge dx^3\nonumber\\&\equiv&\sum_H\varphi(x,H)dx^H.
\label{eq:expansion}
\end{eqnarray}
The convention is that repeated indices are summed over, and no special
significance is attached to whether an index is super- or subscript.
In dealing with quantities defined on the whole space
$\Lambda=\bigoplus_{p=0}^{3}
{^p\Lambda}$, it is convenient to define the main automorphism
\begin{equation}
{\cal A}:\;\Lambda\mapsto\Lambda,\;\;\;{\cal
A}\Phi=\sum_H(-1)^{p(H)}\varphi(x,H)dx^H,
\label{eq:A}
\end{equation}
and the main antiautomorphism
\begin{equation}
{\cal B}:\;\Lambda\mapsto\Lambda,\;\;\;{\cal
B}\Phi=\sum_H(-1)^{_{p(H)}C_2}
\varphi(x,H)dx^H.
\label{eq:B}
\end{equation}
where for $p\in\{0,\ldots,3\}$ the combinatoric factor $_pC_2$ takes values
$\{0,0,1,3\}$.

Three key operations are then:

\begin{itemize}

\item
Exterior derivative 
\begin{equation}
d:\;^p\Lambda\mapsto\,^{p+1}\Lambda,\;\;\;d\Phi=dx^\mu\wedge\partial_\mu\Phi
\label{eq:d}
\end{equation}
Since $\partial_\mu\partial_\nu=\partial_\nu\partial_\mu$ it immediately follows
from (\ref{eq:exterior}) that $d^2=0$. 

\item
Hodge Star
\begin{equation}
\star:\;^p\Lambda\mapsto\,^{3-p}\Lambda,\;\;\;\star dx^H=\rho_{H,{\cal
C}H}dx^{{\cal C}H}
\end{equation}
where ${\cal C}H$ is the complement of $H$.
In odd-dimensional Euclidean spacetimes $\star\star=1$.

\item
Co-derivative
\begin{equation}
\delta:\;^p\Lambda\mapsto\,^{p-1}\Lambda,\;\;\;\delta=\star{\cal
B}\,d\star\!{\cal B}=\star d\!\star\!{\cal A}.
\label{eq:co-deriv}
\end{equation}
and it immediately follows from $d^2=0$, $\star\star=1$ that $\delta^2=0$. 
The co-derivative's sign depends in general on $p$, $d$ and the signature of the
metric~\cite{GS}, which in (\ref{eq:co-deriv}) is captured by the use of the
automorphisms (\ref{eq:A},\ref{eq:B}). A convenient representation for its action is
\begin{equation}
\delta\Phi=-e^\mu\lrcorner\partial_\mu\Phi,
\label{eq:delta}
\end{equation}
where the contraction operator enabling differentiation with
respect to a differential is defined by
\begin{equation}
e^K\lrcorner dx^H=\begin{cases}
\rho_{K,H\backslash K}dx^{H\backslash K} & \mbox{if}\; K\subset H;\\
0 & \mbox{otherwise.}
\end{cases}
\end{equation}

\end{itemize}

\section{The K\"ahler-Dirac Equation}
\label{sec:KDE}

The starting point is the observation that $(d-\delta)^2=-(d\delta+\delta
d)=\partial_\mu\partial_\mu=\Delta$, the Laplacian operator. Hence $d-\delta$ is
in effect the square-root of the Laplacian, and therefore linear in
momentum, while still local. It is thus a candidate for incorporating
in a
relativistic wave equation, as first written by K\"ahler~\cite{Kaehler}:
\begin{equation}
(d-\delta+m)\Phi=0.
\label{eq:KDE}
\end{equation}
The K\"ahler-Dirac equation (KDE) takes the same form in any spacetime dimension.
The scalar parameter $m$ is the fermion mass.
Note that since $d$ and $\delta$ implement $\Delta p=\pm1$, the equation only
makes sense if $\Phi\in\Lambda$, ie. $\Phi$ admits an expansion of the form
(\ref{eq:expansion}), with components $\varphi(x,H)$ having mass dimension 1 in
three spacetime dimensions. 

It is helpful to define the Clifford  product between
differential forms:
\begin{equation}
\vee:\;\Lambda\times\Lambda\mapsto\Lambda,\;\Phi\vee\Xi=\sum_p{{(-1)^{_pC_2}}\over
p!}({\cal A}^pe^{\mu_1}\lrcorner\ldots e^{\mu_p}\lrcorner\Phi)\wedge
(e^{\mu_1}\lrcorner\ldots e^{\mu_p}\lrcorner\Xi),
\end{equation}
with particular instances
\begin{equation}
dx^\mu\vee\Phi=dx^\mu\wedge\Phi+e^\mu\lrcorner\Phi;\;\;\;
\Phi\vee dx^\mu=\Phi\wedge dx^\mu-e^\mu\lrcorner{\cal A}\Phi.
\end{equation}
It immediately follows from (\ref{eq:d}) and (\ref{eq:delta}) that the KDE
can be rewritten
\begin{equation}
(dx^\mu\vee\partial_\mu+m)\Phi=0.
\label{eq:KDEvee}
\end{equation}

Now, the identity 
\begin{equation}
dx^\mu\vee dx^\nu\vee+dx^\nu\vee dx^\mu\vee=2\delta^{\mu\nu}
\label{eq:algebra}
\end{equation}
is strongly reminiscent of the defining relation 
$\{\gamma_\mu,\gamma_\nu\}=2\delta_{\mu\nu}$ for Dirac matrices in
Euclidean metric, and suggests the operation
$dx^\mu\vee$ furnishes a representation of the Dirac algebra in the 8-dimensional
space spanned by $dx^H$. The appropriate representation of the algebra in 3
spacetime dimensions was identified in \cite{Burden:1986by} 
in a study of the staggered lattice fermion operator. It is the direct
sum $\sigma_H\oplus\tau_H$ of two inequivalent irreducible 2-dimensional representations
generated by the Pauli matrices 
$\sigma_\mu$ ($\mu=1,2,3$), and by $\tau_\mu=-\sigma_\mu$. The Pauli matrices
have the property  $\sigma_\mu^*=\sigma_\mu^T$, where * denotes complex
conjugation and $T$ the matrix transpose.
Analysis proceeds by identifying a  new basis
\begin{eqnarray}
\Sigma\oplus
T&=&1+(\sigma_\mu^T\oplus\tau_\mu^T)dx^\mu+{1\over2!}(\sigma_\mu^T\sigma_\nu^T\oplus\tau_\mu^T\tau_\nu^T)dx^\mu\wedge
dx^\nu\label{eq:basis}\\&+&(\sigma_1^T\sigma_2^T\sigma_3^T\oplus\tau_1^T\tau_2^T\tau_3^T)dx^1\wedge
dx^2\wedge dx^3=\sum_H{\cal B}(\sigma\oplus\tau)_H^Tdx^H\nonumber
\end{eqnarray}
The key result is now 
\begin{equation}
dx^\alpha\vee(\Sigma\oplus T)=(\sigma^\alpha\oplus\tau^\alpha)^T(\Sigma\oplus
T)=\sum_c(\sigma^\alpha\oplus\tau^\alpha)^T_{ac}(\Sigma\oplus
T)_{cb},\label{eq:key}
\end{equation}
where Roman indices $a,b,c=1,2$. The derivation of (\ref{eq:key})
makes repeated use of
$\{(\sigma\oplus\tau)_\mu,(\sigma\oplus\tau)_\nu\}=2\delta_{\mu\nu}\otimes\mathbb{1}_{2\times2}$.

In order to express the KDE in the basis (\ref{eq:basis}) we need the orthogonality
relations 
\begin{equation}
\sum_H\sigma_{ab}^H\sigma_{cd}^{H*}=\sum_H\tau_{ab}^H\tau_{cd}^{H*}=4\delta_{ac}\delta_{bd};\;\;\;
\sum_H\sigma_{ab}^H\tau_{cd}^{H*}=0,
\end{equation}
implying
\begin{equation}
dx^H={1\over4}(-1)^{_pC_2}\mbox{tr}(\sigma_H\oplus\tau_H)^{*}(\Sigma\oplus T).
\end{equation}
Using (\ref{eq:expansion}) we then define
\begin{equation}
\Phi(x)=\sum_H\varphi(x,H)dx^H=\sum_{a,b}u_a^b(x)\Sigma_a\oplus d_a^b(x)T_a
\end{equation}
where we have introduced fields $u$, $d$ whose lower index $a=1,2$ will turn out
to be
associated with spinor degrees of freedom in the non-interacting case, and whose upper index $b=1,2$ will be
associated with taste. The field transformations between bases are then:
\begin{eqnarray}
\varphi(x,H)&=&\sum_{a,b}u_a^b(x)(\sigma_H^T)_{ab}\oplus
d_a^b(x)(\tau_H^T)_{ab};\\
(u_a^b\oplus d_a^b)(x)&=&{1\over4}\sum_H\varphi(x,H)(\sigma\oplus\tau)^H_{ab}.
\label{eq:bases}
\end{eqnarray}
Combining the result (\ref{eq:key}) with the KDE
equation (\ref{eq:KDEvee}) we deduce
\begin{equation}
(\gamma_\mu\partial_\mu+m)\psi^b(x)=0,
\end{equation}
ie. the free Dirac equation for a two-taste four-component spinor field $\psi=u\oplus d$,
with Euclidean Dirac matrices defined
\begin{equation}
\gamma_\mu=\left(
\begin{matrix}
\sigma_\mu&\cr 
& \tau_\mu 
\end{matrix}\right).
\end{equation}
We will refer to this familar form as the free KDE in the $\psi$-basis.

\section{Interaction Current}
\label{sec:current}

In order to develop an interacting theory we will need a definition of
a current in the K\"ahler-Dirac formalism. This requires the
definition of a generalised scalar product
$(,)_p:\;\Lambda\times\Lambda\mapsto^{3-p}\!\Lambda$~\cite{Kaehler,Becher:1982ud}. 
The two cases we will need
have $p=0$:
\begin{equation}
(\Phi,\Xi)_0=({\cal B}\Phi\vee\Xi)\wedge\varepsilon
\end{equation}
and $p=1$:
\begin{equation}
(\Phi,\Xi)_1=e_\mu\lrcorner(dx^\mu\vee\Phi,\Xi)_0=e_\mu\lrcorner[(dx^\mu\vee\Phi\vee{\cal
B}\Xi)\wedge\varepsilon],
\label{eq:(,)1}
\end{equation}
where $\varepsilon$ is the volume 3-form $dx^1\wedge dx^2\wedge dx^3$.
In components these are expressed
\begin{equation}
(\Phi,\Xi)_0=\left[\varphi_\emptyset\xi_\emptyset+\varphi_\mu\xi_\mu+{1\over2!}\varphi_{\mu\nu}\xi_{\mu\nu}
+\varphi_{123}\xi_{123}\right]\varepsilon=\sum_H\varphi(x,H)\xi(x,H)\varepsilon;
\end{equation}
and
\begin{eqnarray}
(\Phi,\Xi)_1&=&{1\over2!}\Bigl[\varphi_\emptyset\xi_\alpha+\varphi_\alpha\xi_\emptyset+\varphi_\mu\xi_{\alpha\mu}
+\varphi_{\alpha\mu}\xi_\mu\label{eq:(,)1components}\\
&+&{\textstyle{1\over2!}}(\varphi_{\mu\nu}\xi_{\alpha\mu\nu}+\varphi_{\alpha\mu\nu}\xi_{\mu\nu})\Bigr]
\epsilon_{\alpha\mu\nu}dx^\mu\wedge dx^\nu\nonumber
\end{eqnarray}
The following Green's formula identity is useful~\cite{Kaehler}:
\begin{equation}
d(\Phi,\Xi)_1=(\Phi,(d-\delta)\Xi)_0+((d-\delta)\Phi,\Xi)_0.
\label{eq:identity}
\end{equation}
Next define $\bar\Phi={\cal A}\Phi^*$ as the solution of the adjoint KDE:
\begin{equation}
(d-\delta-m)\bar\Phi=0.
\label{eq:adjointKDE}
\end{equation}
A current 1-form is then given by
\begin{equation}
j=j_\mu dx^\mu=-\star {i\over 4}(\bar\Phi,\Phi)_1.
\label{eq:current}
\end{equation}
Current conservation follows using
(\ref{eq:identity},\ref{eq:KDE},\ref{eq:adjointKDE}):
\begin{eqnarray}
\delta j=-\star d\star j&=&\star
{i\over4}d(\bar\Phi,\Phi)_1=\star {i\over4}[(\bar\Phi,(d-\delta)\Phi)_0+((d-\delta)\bar\Phi,\Phi)_0]\\&=&
-\star {i\over4}(\bar\Phi,\Phi)_0(m-m)=0,\nonumber
\end{eqnarray}
ie. $\partial_\mu j_\mu=0$. Now use $(\Phi,\Xi)_p=(-1)^{_pC_2}(\Xi,\Phi)_p$ and
(\ref{eq:key}) to write
\begin{equation}
(\bar\Phi,\Phi)_1=e_\mu\lrcorner(\bar\Phi,dx^\mu\vee\Phi)_0=e_\mu\lrcorner
(\bar u\Sigma^*\oplus\bar dT^*,(\sigma^\mu
u)\Sigma\oplus(\tau^\mu d)T)_0.
\end{equation}
In the $\psi$-basis the current 1-form
thus reads
\begin{equation}
j=i\sum_b\bar\psi^b(x)\gamma_\mu\psi^b(x)dx^\mu.
\end{equation}

\section{Action and Symmetries}
\label{sec:action}

Now we have enough equipment to define the action and hence the Euclidean path
intgeral. The action for free fields is 
\begin{equation}
S_0={1\over4}\int(\bar\Phi,(d-\delta+m)\Phi)_0
=\sum_{b=1,2}\int\bar\psi^b(\gamma_\mu\partial_\mu+m)\psi^b\varepsilon.\label{eq:S0}
\end{equation}
For the Thirring model this is supplemented by a contact interaction of the form
$-{g^2\over4}j_\mu j_\mu$, where the normalisation of the coupling strength,
which has mass dimension -1, is
somewhat conventional. In the language of forms this reads
\begin{equation}
-{g^2\over4}\int(j,j)_0=-{g^2\over4}\int(j\vee j)\wedge\varepsilon
={g^2\over64}\int(\star(\bar\Phi,\Phi)_1\vee\star(\bar\Phi,\Phi)_1)\wedge\varepsilon.
\end{equation}
Using $(\star\Phi,\star\Xi)_0=(\Phi,\Xi)_0$ we arrive at the Thirring model
action
\begin{eqnarray}
S&=&\int{1\over4}(\bar\Phi,(d-\delta+m)\Phi)_0
+{g^2\over64}((\bar\Phi,\Phi)_1\vee(\bar\Phi,\Phi)_1)\wedge\varepsilon\label{eq:SThir}\\
&=&\int\Bigl[\bar\psi^b(\gamma_\mu\partial_\mu+m)\psi^b
+{g^2\over4}(\bar\psi^b\gamma_\mu\psi^b)(\bar\psi^c\gamma_\mu\psi^c)\Bigr]\varepsilon.
\label{eq:SThirpsi}
\end{eqnarray}

As a consequence of its construction from $(\bar\Phi,\Phi)$ bilinears
the action (\ref{eq:SThir}) has two manifest global symmetries. First:
\begin{equation}
\Phi\mapsto e^{i\theta}\Phi;\;\;\;\bar\Phi\mapsto e^{-i\theta}\bar\Phi.
\label{eq:rot}
\end{equation}
This symmetry correponds to the conservation of fermion charge, and the
corresponding Noether current is given by (\ref{eq:current}). Second, in the
limit $m\to0$:
\begin{equation}
\Phi\mapsto e^{i\omega{\cal A}}\Phi;\;\;\;\bar\Phi\mapsto e^{i\omega{\cal
A}}\bar\Phi,
\label{eq:rotA}
\end{equation}
which follows because $d,\delta$ both yield $\Delta p=\pm1$, and by inspection
of the component expansion  of $(\bar\Phi,\Phi)_1$ (\ref{eq:(,)1components}). This is
analogous to the chiral symmetry protecting fermions from additive mass
renormalisation in $d=4$. The corresponding Noether current is
\begin{equation}
j_{\cal A}=-\star {i\over 4}(\bar\Phi,{\cal A}\Phi)_1.
\label{eq:currentAPhi}
\end{equation}
In order to translate to the $\psi$-basis, observe that the action of ${\cal A}$
in effect exchanges $\sigma_H$ and $\tau_H$ in (\ref{eq:bases}). It then follows
straightforwardly that 
\begin{equation}
j_{{\cal A}\mu}=i\bar\psi^b\gamma_\mu\gamma_5\psi^b
\label{eq:currentA}
\end{equation}
where we introduce two new hermitian $\gamma$-matrices obeying
$\{\gamma_4,\gamma_\mu\}=\{\gamma_5,\gamma_\mu\}=\{\gamma_4,\gamma_5\}=0$:
\begin{equation}
\gamma_4=\left(
\begin{matrix}
&-i\mathbb{1}\cr 
i\mathbb{1}&  
\end{matrix}\right);\;\;\;
\gamma_5=\left(
\begin{matrix}
&\mathbb{1}\cr 
\mathbb{1}&  
\end{matrix}\right).
\end{equation}

From here it is straightforward to extend the model by introducing $N$
K\"ahler-Dirac fermion flavors $\Phi^i$, $i=1,\ldots,N$. The flavor index $i$ is
distinct from the indices $b,c=1,2$ in (\ref{eq:SThirpsi}), which run over 
taste degrees of freedom. 
The two U(1) rotation symmetries (\ref{eq:rot},\ref{eq:rotA}) are
trivially extended to U$(N)\otimes$U$_{\cal A}(N)$, broken to U($N$)
either explicitly by $m\not=0$, or spontaneously by dynamical generation of a
non-vanishing condensate $\langle(\bar\Phi^i,\Phi^i)_0\rangle$.

Finally, consider discrete parity inversion. In odd spacetime dimensions this is
conveniently represented by inversion of all spacetime axes:
$x_\mu\mapsto-x_\mu$, $\partial_\mu\mapsto-\partial_\mu$. The action
(\ref{eq:SThir},\ref{eq:SThirpsi}) is invariant provided
\begin{equation}
\Phi(x)\mapsto{\cal A}\Phi(-x);\;\bar\Phi(x)\mapsto{\cal A}\bar\Phi(-x)
\Rightarrow
\psi(x)\mapsto\gamma_5\psi(-x);\;\bar\psi(x)\mapsto\bar\psi(-x)\gamma_5.
\label{eq:parity}
\end{equation}
Note that the Noether currents (\ref{eq:current},\ref{eq:currentA}), along with all
bilinears of the form $(\Phi,\Xi)_1$, are parity-odd.

The Euclidean path integral is defined by
\begin{equation}
{\cal Z}=\int D\Phi D\bar\Phi\exp(-S[\Phi,\bar\Phi])
\end{equation}
where $\Phi,\bar\Phi$ are now Grassmann-valued and $\bar\Phi$ is considered
independent of $\Phi$. We illustrate its use via a derivation of the Ward
Identity for the divergence of the current $j_{\cal A}$; for simplicity we
consider only the free action (\ref{eq:S0}). Consider the impact of the
field transformation (\ref{eq:rotA}) where $\omega(x)$ is infinitesimal but
now spacetime-dependent.
\begin{eqnarray}
S_0\mapsto S_0^\prime&=&{1\over4}\int\left(e^{i\omega(x){\cal
A}}\bar\Phi,(d-\delta+m)e^{i\omega(x){\cal A}}\Phi\right)_0\nonumber\\
&=&{1\over4}\int\left(e^{i\omega{\cal A}}\bar\Phi,dx^\mu\vee\partial_\mu
e^{i\omega{\cal A}}\Phi+me^{i\omega{\cal A}}\Phi\right)_0\\
&=&S_0+{1\over4}\int i\partial_\mu\omega\left(\bar\Phi,dx^\mu\vee{\cal
A}\Phi\right)_0+im\omega\left((\bar\Phi,{\cal A}\Phi)_0+({\cal
A}\bar\Phi,\Phi)_0\right).\nonumber
\end{eqnarray}
Now use (\ref{eq:(,)1})
together with $(\Phi,\Xi)_p=(-1)^{_pC_2}(\Xi,\Phi)_p$ and the definition (\ref{eq:currentAPhi}) to write
\begin{eqnarray}
S_0^\prime-S_0&=&-\int(\partial_\mu\omega)dx^\mu\wedge\star j_{\cal
A}+i{m\over2}\int\omega(\bar\Phi,{\cal A}\Phi)_0\\
&=&\int\omega\left(d \star j_{\cal A}+i{m\over2}(\bar\Phi,{\cal
A}\Phi)_0\right),\nonumber
\end{eqnarray}
where in the second step we have integrated the first term by parts. Since the
path-integral measure $D\Phi D\bar\Phi=\prod_{x,H}d\varphi(x,H)d\bar\varphi(x,H)$ 
is formally invariant under the field
transformation, the change of variables has no impact on the path integral, and
we conclude
\begin{equation}
\left\langle\int\omega\left[-\star\delta j_{\cal A}+i{m\over2}(\bar\Phi,{\cal
A}\Phi)_0\right]\right\rangle=0.
\label{eq:WI}
\end{equation}
Since (\ref{eq:WI}) holds for any $\omega(x)$, we conclude the expectation value
of the 3-form in square brackets is identically zero, which is the Ward
Identity. In the $\psi$-basis it has the familiar form
\begin{equation}
\left\langle\partial_\mu\bar\psi\gamma_\mu\gamma_5\psi-2m\bar\psi\gamma_5\psi\right\rangle=0.
\end{equation}

\section{Impact of Quantum Corrections}
\label{sec:qc}

Our treatment up to this point has been either classical or formal. In any
application to a genuine interacting quantum field theory, it is inevitable that
the theory will need to be regularised somehow in order to control the calculation of
quantum corrections. As a concrete example, we have already discussed the close
parallels between the KDE continuum formalism and staggered lattice fermions,
and will assume without further discussion that the proof of
\cite{Becher:1982ud} that the KDE is the formal continuum limit of staggered
fermions continues to apply in 3 dimensions. 

Regularisation is essentially some kind of truncation of the
degrees of freedom present in the classical field theory, and inevitably
violates some of the symmetries of the classical theory. In many cases
this leads to the requirement of renormalisation of both the fields and the coupling
parameters in the theory, which depends on some physical scale. 
As a concrete example, consider the Thirring action
in the $\psi$-basis (\ref{eq:SThirpsi}), where the rotations 
(\ref{eq:rot},\ref{eq:rotA}) take the form 
\begin{equation}
\psi\mapsto e^{i\theta}\psi;\;\;\;\bar\psi\mapsto\bar\psi
e^{-i\theta}:\;\;\;\;\;
\psi\mapsto e^{i\theta\gamma_5}\psi;\;\;\;\bar\psi\mapsto\bar\psi
e^{i\theta\gamma_5}
\label{eq:rotrotApsi}
\end{equation}
Eqn.~(\ref{eq:SThirpsi}) also
looks to be invariant under a
U(2) rotation among the tastes indexed by $b,c$. Beyond that, in the limit
$m\to0$ there is an additional symmetry 
\begin{equation}
\psi\mapsto e^{i\theta\gamma_4}\psi;\;\;\;\bar\psi\mapsto\bar\psi
e^{i\theta\gamma_4}
\label{eq:gamma4}
\end{equation}
as well as 
\begin{equation}
\psi\mapsto e^{\theta\gamma_4\gamma_5}\psi;\;\;\;\bar\psi\mapsto\bar\psi
e^{-\theta\gamma_4\gamma_5},
\label{eq:gamma45}
\end{equation}
valid for any $m$. Rotations (\ref{eq:gamma4},\ref{eq:gamma45}) combined  with
(\ref{eq:rotrotApsi}) and 
the taste rotations would generate a U(4$N$) global symmetry broken to
U($2N)\otimes$U($2N)$ by a fermion mass. Our viewpoint is that this symmetry is
not fundamental and can only be recovered in certain limits, such as long
wavelength or weak coupling. 

We will proceed on the assumption that the geometric description employed in the KDE
is more natural, so that after quantum corrections the field expansion of
eqn.~(\ref{eq:expansion}) is modified:
\begin{equation}
\Phi_r(x)=\sum_H Z_{p(H)}\varphi(x,H)dx^H.
\label{eq:expansionR}
\end{equation}
Here a renormalised field $\Phi_r$ is defined in terms of bare components
$\varphi(x,H)$ via wavefunction renormalisation constants $Z_p$ which depend on
the interaction strength, the renormalisation scale and, crucially in this
context, on the form degree $p$. This correction is covariant, in the sense
that $Z_p$ is insensitive to rotations acting on the spacetime indices specific to
$\varphi(x,H)$, and the key symmetries  (\ref{eq:rot}) and
(\ref{eq:rotA}) continue to be respected by $\Phi_r$ even with $Z_p\not=1$.

The form of (\ref{eq:expansionR}) motivates a more general exploration of
possible interaction currrents. In $d=3$ the space of bilinear currents
consistent with the four renormalisation constants $Z_p$ is spanned by 
$\star(\bar\Phi,\Phi)_1$, $\star(\bar\Phi,{\cal A}\Phi)_1$, $\star({\cal B}\bar\Phi,\Phi)_1$ and 
$\star({\cal B}\bar\Phi,{\cal A}\Phi)_1$. Transcription to the $\psi$-basis for the
first two of these is given in (\ref{eq:current},\ref{eq:currentA}), and eg.
\begin{eqnarray}
({\cal B}\bar\Phi,\Phi)_1&=&e_\mu\lrcorner({\cal B}\bar\Phi,dx^\mu\vee\Phi)_0\\
&=&e_\mu\lrcorner\varepsilon\sum_H(-1)^{_pC_2}\bigl[\bar u\sigma_H\oplus\bar
d\tau_H\bigr]\bigl[(\sigma_\mu u)\sigma_H^*\oplus(\tau_\mu
d)\tau_H^*\bigr]\nonumber
\end{eqnarray}
Now observe the following identities for the components of $\sigma_H$:
\begin{eqnarray}
\sum_\rho\sigma_\rho\mathbb{1}\sigma_\rho&=&3;\;\;\;
\sum_\rho\sigma_\rho\sigma_1\sigma_2\sigma_3\sigma_\rho=3\sigma_1\sigma_2\sigma_3;\nonumber\\
\sum_\rho\sigma_\rho\sigma_\mu\sigma_\rho&=&-\sigma_\mu;\;\;
\sum_\rho\sigma_\rho\sigma_\mu\sigma_\nu\sigma_\rho=-\sigma_\mu\sigma_\nu.
\end{eqnarray}
Recalling $\tau_H=(-1)^p\sigma_H$, we deduce a particularly convenient combination:
\begin{eqnarray}
-\star(\bar\Phi\!\!\!&+&\!\!\!\!2{\cal B}\bar\Phi,\Phi)_1\nonumber\\
&=&dx^\mu
\sum_H\sum_\rho\bigl[\bar u\sigma_H\oplus\bar d\tau_H\bigr]
\bigl[(\tau_\rho\tau_\mu d\tau_\rho^*)^b_a(\sigma_H^*)_{ab}\oplus
(\sigma_\rho\sigma_\mu u\sigma_\rho^*)^b_a(\tau_H^*)_{ab}\bigr]\nonumber\\
&=&4\sum_\rho(\bar u,\bar d)\left(\begin{matrix}
&-\sigma_\rho\sigma_\mu\otimes\sigma_\rho^*\\
+\sigma_\rho\sigma_\mu\otimes\sigma_\rho^*&\end{matrix}\right)\left(\begin{matrix}u\\d\end{matrix}\right)
dx^\mu\\
&=&4\sum_\rho\bar\psi
(i\gamma_4\gamma_\rho\gamma_\mu\otimes\tau_\rho^*)\psi dx^\mu.\nonumber
\end{eqnarray}
Here the second component of the tensor product is a $2\times2$ matrix acting on
taste indices.  Similarly,
\begin{equation}
-\star(\bar\Phi+2{\cal B}\bar\Phi,{\cal A}\Phi)_1
=-4\sum_\rho\bar\psi
(i\gamma_4\gamma_5\gamma_\rho\gamma_\mu\otimes\tau_\rho^*)\psi dx^\mu.
\end{equation}

In either case what emerges is an interaction current which although parity-odd and respecting the
U(1)$\otimes$U$_{\cal A}$(1) symmetries (\ref{eq:rot},\ref{eq:rotA}) no longer treats fermion tastes as
independent degrees of freedom but rather entangles taste and
spacetime rotations, contrary to what is expected for particle flavor degrees
of freedom. Remarkably, the currents $\star(\bar\Phi,\Phi_1)$,
$\star(\bar\Phi,{\cal A}\Phi)_1$, $\star([1+2{\cal B}]\bar\Phi,\Phi)_1$ and
$\star([1+2{\cal B}]\bar\Phi,{\cal A}\Phi)_1$ all feature in equal weight contact
interactions in the Thirring model formulated with staggered fermions on a
3$d$ cubic lattice 
as derived in a basis with explicit spinor 
and taste indices using the formalism of \cite{Burden:1986by}, 
and given in Eqn.~(2.12) of \cite{DelDebbio:1997dv}. 
In view of the equivalence~\cite{Becher:1982ud} between  K\"ahler-Dirac fermions
and the
formal continuum limit of
staggered lattice fermions, this result is not surprising. 

It is now clear the 
interactions survive the long-wavelength $a\to0$ limit, where the lattice
spacing $a$ furnishes an explicit UV cutoff. Other terms entangling spinor and taste
degrees of freedom which formally vanish as $O(a)$ are also
present in the lattice formulation~\cite{DelDebbio:1997dv}. The current analysis demonstrates that
spin/taste entanglement is not a lattice artifact, but is rooted in a continuum
action of the form (\ref{eq:SThir}) with U($N)\otimes$U$_{\cal A}(N)$
symmetry. However, it is significant that such terms also emerge from a well-defined
regularisation capable of exploring strongly-interacting dynamics. 

\section{Reduced K\"ahler-Dirac Fermions}
\label{sec:trunc}

K\"ahler-Dirac fermions offer a new language with which to discuss relativistic fermion
dynamics. 
To quote Becher and
Joos, ``This differential geometric description of fermions might be a basis for
the construction of different finds of field theoretic
model,''~\cite{Becher:1982ud}. 
Once the differential geometric scaffolding has been removed, 
what kind of stories will we be able to tell? 
With motivation
coming from a desire to understand novel structures at strongly-interacting
fixed points, in this section we will hazard some speculations.

Let's start by expressing the free action (\ref{eq:S0}) in the $\varphi$-basis
(\ref{eq:expansion}),
with Lagrangian density
\begin{eqnarray}
\star{1\over4}(\bar\Phi,(d-\delta)\Phi)_0&=&{1\over4}\biggl[\bar\varphi_\emptyset\partial_\mu\varphi_\mu+
\bar\varphi_\mu(\partial_\mu\varphi_\emptyset+\partial_\nu\varphi_{\nu\mu})\nonumber\\
&+&\bar\varphi_{\mu\nu}\left(\partial_\mu\varphi_\nu+{1\over2!}\partial_\lambda\varphi_{\lambda\mu\nu}\right)
+{1\over2!}\bar\varphi_{\mu\nu\lambda}\partial_\mu\varphi_{\nu\lambda}\biggr].\label{eq:KDEphi}
\end{eqnarray}
Each term in (\ref{eq:KDEphi}) is separately invariant under
U(1)$\otimes$U$_{\cal A}(1)$ and parity (\ref{eq:parity}).
Now consider a reduced action containing just a subset of the $p$-form
fields $\varphi(x,H)$. The motivation comes from Eqn.~(\ref{eq:expansionR}), where we envisage
a partition of $\{0,1,2,3\}$ into sets $P,Q$ with $Z_{p\in P}\gg Z_{p\in
Q}\approx0$ arising,
say, as a consequence of large anomalous scaling dimensions at a
renormalisation group fixed point. Clearly only cases 
retaining consecutive values of $p$ will result in propagating states. We consider two examples.

\subsection{$P=\{0,1\}$}

If we truncate the field content to just $p=0,1$ there are four components
$\phi\equiv(\varphi_\emptyset,\varphi_\mu)^T$ to keep track of. 
The Lagrangian density is 
\begin{equation}
{\cal
L}_{01}={1\over4}[\bar\varphi_\emptyset\partial_\mu\varphi_\mu+\bar\varphi_\mu\partial_\mu\varphi_\emptyset
+m\bar\varphi_\emptyset\varphi_\emptyset+m\bar\varphi_\mu\varphi_\mu]\equiv{1\over4}\bar\phi
M\phi.\label{eq:L01}
\end{equation}
The $4\times4$
matrix $M$ has
\begin{equation}
\mbox{det}M=m^2(m^2-\Delta),
\end{equation}
which therefore vanishes identically for massless fermions. The
propagator $M^{-1}$ has components
\begin{eqnarray}
\langle\varphi_\emptyset\bar\varphi_\emptyset\rangle&=&{m\over{m^2-\Delta}};\;\;
\langle\varphi_\mu\bar\varphi_\nu\rangle={m\delta_{\mu\nu}\over{m^2-\Delta}}-{{{\cal
P}_{\mu\nu}\Delta}\over{m(m^2-\Delta)}};\nonumber\\
\langle\varphi_\emptyset\bar\varphi_\mu\rangle&=&\langle\varphi_\mu\bar\varphi_\emptyset\rangle=
-{\partial_\mu\over{m^2-\Delta}},
\end{eqnarray}
where the transverse projector
\begin{equation}
{\cal P}_{\mu\nu}=\delta_{\mu\nu}-{\partial_\mu\partial_\nu\over\Delta}
\end{equation}
such that ${\cal P}_{\mu\nu}\partial_\mu=0$, ${\cal P}_{\mu\lambda}{\cal
P}_{\lambda\nu}={\cal P}_{\mu\nu}$, and $\mbox{tr}{\cal P}=2$. In momentum space, all
components manifest a particle pole at $k^2=-m^2$, but asymptotically scale
differently: $\emptyset\emptyset\sim k^{-2}$; $\mu\nu\sim k^{-2}+k^0$;
$\emptyset\mu\sim k^{-1}$.
We conclude that ${\cal L}_{01}$ describes particles of mass $m$, and that the
resulting effective theory is well-behaved in the IR regime $k^2\lesssim m^2$.
The singular part in the $m\to0$ limit has  vanishing longitudinal component.

Note also that following a 
field redefinition $\varphi_{\mu\nu}=\epsilon_{\mu\nu\lambda}\xi_\lambda$;
$\varphi_{\mu\nu\lambda}=\epsilon_{\mu\nu\lambda}\xi_\emptyset$, 
the action ${\cal L}_{23}[\xi_{\emptyset,\mu},\bar\xi_{\emptyset,\mu}]$ 
yields an action identical in form to (\ref{eq:L01}), 
the only
difference being that in contrast to (\ref{eq:L01}) the field $\xi_\emptyset$
has negative intrinsic parity and
$\xi_\mu$ positive.

\subsection{$P=\{1,2\}$}

In this case there are 6 field components $\varphi_\mu,\varphi_{\mu\nu}$ with
Lagrangian density 
\begin{equation}
{\cal
L}_{12}={1\over4}\left[\bar\varphi_\mu\partial_\nu\varphi_{\nu\mu}+\bar\varphi_{\mu\nu}\partial_\mu\varphi_\nu
+m\bar\varphi_\mu\varphi_\mu+{m\over2!}\bar\varphi_{\mu\nu}\varphi_{\mu\nu}\right].
\end{equation}
After a field redefinition
\begin{equation}
\chi_\mu=\varphi_\mu;\;\;\;\bar\chi_\mu={1\over2!}\epsilon_{\mu\nu\lambda}\bar\varphi_{\nu\lambda};\;\;
\bar\xi_\mu=\bar\varphi_\mu;\;\;\;\xi_\mu={1\over2!}\epsilon_{\mu\nu\lambda}\varphi_{\nu\lambda},
\end{equation}
${\cal L}_{12}$ can be rewritten
\begin{eqnarray}
{\cal
L}_{12}&=&{1\over4}\Bigl[\epsilon_{\mu\nu\lambda}\left[\bar\chi_\mu\partial_\nu\chi_\lambda-
\bar\xi_\mu\partial_\nu\xi_\lambda\right]+m(\bar\chi_\mu\xi_\mu+\bar\xi_\mu\chi_\mu)\Bigr]\label{eq:CS}\\
&=&{1\over4}(\bar\chi,\bar\xi)\left(\begin{matrix}
\partial_\mu\lambda^\mu&m\mathbb{1}_{3\times3}\\m\mathbb{1}_{3\times3}&-\partial_\mu\lambda^\mu
\end{matrix}\right)\left(\begin{matrix}\chi\\
\xi\end{matrix}\right)\equiv{1\over4}\bar\Upsilon M\Upsilon,\label{eq:spin1}
\end{eqnarray}
where we have implicitly defined 6 component fermion fields
$\Upsilon,\bar\Upsilon$.
The individual kinetic terms for the three-component objects $\chi,\xi$ 
in (\ref{eq:CS}) superficially resemble the Chern-Simons action for
gauge boson fields in $d=3$, and in (\ref{eq:spin1}) the $3\times3$ matrices
$\lambda^\mu$, $\mu=1,\ldots,3$ are antihermitian generators of the spin-1 representation of
SU(2), ie. obeying
$[\lambda_\mu,\lambda_\nu]=-\epsilon_{\mu\nu\rho}\lambda_\rho$, and each with
eigenvalues $i\lambda=0,\pm1$. Fields respond to rotations in the $\rho\sigma$
plane via $(\chi,\xi)^T_\mu\mapsto(\Lambda_1)_{\mu\nu}(\chi,\xi)^T_\nu$  with 
\begin{equation}
\Lambda_1(\theta_{\rho\sigma})=\exp\left(-{{\theta_{\rho\sigma}}\over2!}[\lambda^\rho,\lambda^\sigma]\right).
\end{equation}

Unlike Dirac matrices the $\lambda^\mu$ don't obey a Clifford algebra,
so the $6\times6$ matrix $M$ in (\ref{eq:spin1}) is less straightforward to invert than a
conventional Dirac operator.  We
start by checking its determinant, introducing the notation
$\partial_\mu\lambda^\mu\equiv\partial\cdot\lambda$:
\begin{eqnarray}
\mbox{det}M&=&-\mbox{det}\left(m^2\mathbb{1}_{3\times3}+(\partial\cdot\lambda)^2\right)\\
&=&-\exp\left[\mbox{tr}(\ln m^2)+\mbox{tr}\left(+{{(\partial\cdot\lambda)^2}\over m^2}
-{{(\partial\cdot\lambda)^4}\over 2m^4}
+{{(\partial\cdot\lambda)^6}\over 3m^6}-\cdots\right)\right]\nonumber
\end{eqnarray}
Now use $(\partial\cdot\lambda)^2=-\Delta{\cal P}$ and
$\mbox{tr}{\cal P}=2$
to write
\begin{equation}
\mbox{det}M=-\exp\left(3\ln m^2+2\ln(1-{\Delta\over
m^2})\right)=-m^2(m^2-\Delta)^2.
\end{equation}
Again, the determinant vanishes if $m=0$. The propagator exists for $m\not=0$
and is given by
\begin{equation}
\langle\Upsilon\bar\Upsilon\rangle={1\over{m^2-\Delta}}\left(\begin{matrix}
\partial\cdot\lambda&m-m^{-1}\partial_\mu\partial_\nu\\
m-m^{-1}\partial_\mu\partial_\nu&-\partial\cdot\lambda\end{matrix}\right).
\end{equation}
Also note that
\begin{eqnarray}
{-1\over{\Delta(m^2-\Delta)}}\left[\begin{matrix}
(\partial\cdot\lambda)^3&
m(\partial\cdot\lambda)^2\\m(\partial\cdot\lambda)^2&
-(\partial\cdot\lambda)^3\end{matrix}\right]M
&=&{1\over{m^2-\Delta}}\left[\begin{matrix}
\partial\cdot\lambda&m\\m&-\partial\cdot\lambda\end{matrix}\right]
{\cal P}M\nonumber\\
&=&{\cal P}_{\mu\nu}\otimes\mathbb{1}_{2\times2}
\end{eqnarray}

The fact that $M$ in the massless limit is invertible when acting on a transverse subspace is
reminiscent of gauge theories, where the same issue occurs due to the redundancy
of the field description as a consequence of an underlying invariance of the
action under local gauge transformations of the form $A_\mu\mapsto
A_\mu+\partial_\mu\Lambda$. We can trace this to the invariance of
(\ref{eq:CS}),
after integration by parts,
under 
\begin{equation}
\chi_\mu\mapsto\chi_\mu+\partial_\mu\vartheta_\chi;\;\;
\bar\chi_\mu\mapsto\bar\chi_\mu+\partial_\mu\vartheta_{\bar\chi};\;\;
\xi_\mu\mapsto\xi_\mu+\partial_\mu\vartheta_\xi;\;\;
\bar\xi_\mu\mapsto\bar\xi_\mu+\partial_\mu\vartheta_{\bar\chi}.
\label{eq:gauge}
\end{equation}
Here the $\vartheta(x)$ are Grassmann-valued  fields, and the subscripts
emphasise that independent shifts are applied to each fermi field. For this
reason the mass term is not in general 
invariant under (\ref{eq:gauge}), consistent with the fact that $M$ is
invertible once  $m\not=0$.  Further note that the textbook solution to defining
a gauge-field propagator, namely to fix a gauge by adding a covariant term of
the form $\zeta^{-1}(\partial_\mu A_\mu)^2$ to the action, would in this case 
yield terms of the form , eg.
$\sim(\epsilon_{\mu\nu\lambda}\partial_\mu\bar\varphi_{\nu\lambda})(\partial_\rho\varphi_\rho)$,
consistent with U(1)$_{\cal A}$ but violating  parity. Rather, it makes more
sense to regard the term $m(\bar\chi\xi+\bar\xi\chi)$ as the ``gauge-fixing
term''.

We conclude that ${\cal L}_{12}$ describes a fermion field transforming in the
spin-1 representation of the rotation group, with some features reminiscent of a
gauge field, namely that in the UV limit the only remaining degrees of freedom
are transverse, ie. helicity eigenstates, so that six components are reduced to
four.  The Noether currents corresponding to symmetries
(\ref{eq:rot},\ref{eq:rotA}) are given by 
\begin{equation}
j_\mu,\,j_{{\cal
A}\mu}=-{i\over4}\left[\bar\chi\lambda_\mu\chi\mp\bar\xi\lambda_\mu\xi\right].
\label{eq:rcurrents}
\end{equation}
Assigning 3 as the timelike direction, we identify a fermion charge operator
$-i\lambda^3\otimes\sigma_3$ with $\pm$ restframe eigenstates $F,\bar F=(1,\mp
i,0,1,\pm
i,0)^T$, ie. fermions 
(antifermions)  correspond to left(right)- and right(left)-handed circularly
polarised $\chi$($\xi$)-states, 
which remain transverse under SO(3) rotations.
Asymptotically the propagator scales as $k^{-1}$, as expected for a
relativistic fermion. The propagator pole 
at $k^2=-m^2$ again corresponds to a physical particle.

Finally, we can use the fermion current of (\ref{eq:rcurrents}) to write the
Lagrangian for the Thirring model based on ${\cal L}_{12}$, using the
$\Upsilon$-basis:
\begin{equation}
{\cal L}_{\rm
rThir}={1\over4}\bar\Upsilon(\partial\cdot\lambda\otimes\sigma_3+m\mathbb{1}\otimes\sigma_1)\Upsilon
+{g^2\over64}\left(\bar\Upsilon\lambda_\mu\otimes\sigma_3\Upsilon\right)^2.
\label{eq:spin1fermions}
\end{equation}
In the same basis the invariances (\ref{eq:rot},\ref{eq:rotA}) read
\begin{eqnarray}
\Upsilon\mapsto e^{i\alpha}\Upsilon&;&\;\;\;\bar\Upsilon\mapsto\bar\Upsilon e^{-i\alpha}\\
\Upsilon\mapsto
e^{i(\mathbb{1}\otimes\sigma_3)\alpha}\Upsilon&;&\;\;\;\bar\Upsilon\mapsto\bar\Upsilon
e^{-i(\mathbb{1}\otimes\sigma_3)\alpha}
\end{eqnarray}
while parity is 
\begin{equation}
\Upsilon(x)\mapsto-(\mathbb{1}\otimes\sigma_3)\Upsilon(-x);\;\;\;
\bar\Upsilon(x)\mapsto\bar\Upsilon(-x)(\mathbb{1}\otimes\sigma_3).
\end{equation}

\section{Discussion}
\label{sec:disc}

This paper has developed the description of relativistic fermions 
in the language of differential geometry,
originally set out in \cite{Kaehler}, to three spacetime dimensions. The
principal result is the specification of a continuum field
theory sharing the same parity and global U($N)\otimes$U($N$) invariances 
as the ``staggered Thirring model'' originally studied numerically using lattice field
theory simulations in
\cite{DelDebbio:1997dv}. In our view this puts the staggered Thirring model on
a firm footing as an interacting quantum field theory distinct from the 
U($2N$)-invariant version based on the continuum action (\ref{eq:contThir}),
which is the focus of much recent numerical work.
This result is entirely consistent with Becher and Joos'
demonstration that K\"ahler-Dirac fermions are the correct continuum limit for
staggered lattice fermions~\cite{Becher:1982ud}. Beyond the
weak-coupling and long-wavelength limits, 
we've seen that spin/taste
entanglement is not merely a lattice artifact, but a genuine feature of an
interacting continuum field theory: tastes are not the same as flavors. 

An important consequence of regarding the $\varphi$-basis as more
fundamental than the more familiar $\psi$-basis is the response to
quantum corrections encapsulated in the proposed relation (\ref{eq:expansionR})
relating renormalised to bare fields, in which multiplicative
renormlisation depends solely on $p$, consistent with U($N)\otimes$U($N$)
symmetry. This was demonstrated  explicitly in Sec.~\ref{sec:qc} through the
recovery of interaction currents 
entangling spin and taste originally found in the staggered Thirring model. 
However, a more
spectacular, if speculative consequence  was worked out in Sec.~\ref{sec:trunc},
where the assumption of a strong hierarchy of the $Z_p$ arising due to large
anomalous scaling dimensions in the vicinity of a renormalisation-group fixed
point motivated the investigation of truncated actions retaining just two
$p$-values. In particular the Lagrangian ${\cal L}_{\rm rThir}$
(\ref{eq:spin1fermions}) was found to be particularly compelling, describing
six-component spin-1 fermions, with fermions/antifermions being states of
well-defined polarisation, and dynamics dominated by the four components lying in the
transverse subspace in the UV limit.  Could this exotica form the basis for a
description of strongly interacting fixed-point
dynamics? The answer must await a controlled non-perturbative investigation. 
 
We conclude with a brief discussion of spin and statistics.
The Lagrangian (\ref{eq:spin1fermions}) describes spin-1 fermions 
which in the
canonical approach to field quantisation would be represented by field operators
with anticommutator $\{\Upsilon_\alpha(\vec x,t),\Upsilon_\beta^\dagger(\vec
x^\prime,t)\}=\delta^2(\vec x-\vec x^\prime)\delta_{\alpha\beta}$. An immediate
concern is the apparent contradiction with the spin-statistics theorem requiring 
Lorentz-invariant theories of anti-commuting fields to be quantised with
half-integer spin representations of the Lorentz group. A symptom of the
problem is revealed through the ground state expectation of the anticommutator 
of fields at arbitrary spacetime separation~\cite{B&D}:
\begin{equation}
\langle0\vert\{\Upsilon(x),\bar\Upsilon(x^\prime)\}\vert0\rangle=
i(i\partial\cdot\tilde\lambda\otimes\sigma_3+m\mathbb{1}\otimes\sigma_1)\Delta_{\rm
sym}(x^\prime-x).
\label{eq:anticom}
\end{equation}
Here $\tilde\lambda_\mu$ are Minkowski space versions of the $\lambda$-matrices, 
we have assumed that all states are defined in the transverse subspace, and
for field quantisation with the ``wrong'' statistics  the $PCT$ theorem dictates
the appearance on the RHS of the symmetric solution of the Klein-Gordon equation
(or its generalisation):
\begin{equation}
\Delta_{\rm sym}(x)=\int{{d^2\vec k}\over{(2\pi)^2}}{\cos(k\cdot
x)\over\omega(k)},
\end{equation}
where for free fields $\omega(k)=\sqrt{k^2+m^2}$. Now specialise to the case of 
a spacelike interval $\vec x=x\hat x$ with $\vert\hat x\vert=1$. We find
\begin{equation}
\langle0\vert\{\Upsilon(0),\bar\Upsilon(\vec x)\}\vert0\rangle=
{im^{3\over2}\over{(2\pi x)^{1\over2}}}{1\over\pi}\left[
(i\hat
x\cdot\tilde\lambda\otimes\sigma_3)K_{3\over2}(mx)+(\mathbb{1}\otimes\sigma_1)K_{1\over2}(mx)\right].
\label{eq:Bessel}
\end{equation}
The non-vanishing of the RHS of
(\ref{eq:anticom}) outside the lightcone signals a violation of
microcausality. This is a general result independent of the detailed form of the
dispersion $\omega(k)$. For free fields the asymptotic
properties of the modified Bessel functions in (\ref{eq:Bessel})  can be used to to find
\begin{equation}
\lim_{x\to\infty}\langle0\vert\{\Upsilon(0),\bar\Upsilon(\vec x))\}\vert0\rangle=
{i\over{2\pi}}{m\over x}e^{-mx}[i\hat
x\cdot\tilde\lambda\otimes\sigma_3+\mathbb{1}\otimes\sigma_1],
\end{equation}
and
\begin{equation}
\lim_{x\to0}\langle0\vert\{\Upsilon(0),\bar\Upsilon(\vec x))\}\vert0\rangle=
{i\over{2\pi x^2}}(i\hat
x\cdot\tilde\lambda\otimes\sigma_3)+{im\over{2\pi x}}(\mathbb{1}\otimes\sigma_1);
\end{equation}
that is, the causality violation is localised to within roughly a Compton wavelength of
the lightcone, but diverges as $x\to0$, although less severely than the $x^{-3}$
behaviour of 3+1$d$~\cite{B&D}.

Since microcausality is a desirable property for a fundamental theory, the 
correct relation between spin and statistics is a necessary ingredient of a
complete
quantum field theory. By hypothesis, however, the spin-1 action
(\ref{eq:spin1fermions}) serves only as an effective description of the dynamics near
a UV fixed point, in the deep Euclidean regime $k^2\to\infty$ very far
from the lightcone. The question of whether the spin-statistics linkage
compromises the fixed-point description remains open.

\section*{Acknowledgements}
This work was supported by STFC Consolidated Grant ST/T000813/1.

\end{document}